\documentclass[%
10pt,
twocolumn,
superscriptaddress,
longbibliography
 amsmath,amssymb,
 aps, %physrev,
prb,
floatfix,
]{revtex4-2}

\usepackage{graphicx}% Include figure files
\usepackage{dcolumn}% Align table columns on decimal point
\usepackage{bm}% bold math
\usepackage{xcolor}
\usepackage{physics}
\usepackage{hyperref}% add hypertext capabilities
\usepackage[color=orange!60,textsize=scriptsize]{todonotes}
\usepackage{siunitx}
\DeclareSIUnit\angstrom{\text{Å}}

\begin{document}

% \preprint{}

\title{\textbf{Combining Harmonic Sampling with the Worm Algorithm to Improve the Efficiency of Path Integral Monte Carlo}} 

\author{Sourav Karmakar}
\affiliation{School of Chemistry, Tel Aviv University, Tel Aviv 6997801, Israel.}
\affiliation{Center for Computational Molecular and Materials Science, Tel Aviv University, Tel Aviv 6997801, Israel.}
\author{Sutirtha Paul}
\affiliation{Department of Physics and Astronomy, University of Tennessee, Knoxville, Tennessee 37996, USA.}
\author{Adrian Del Maestro}
\affiliation{Department of Physics and Astronomy, University of Tennessee, Knoxville, Tennessee 37996, USA.}
\affiliation{Min H. Kao Department of Electrical Engineering and Computer Science, University of Tennessee, Knoxville, TN 37996, USA.}
\author{Barak Hirshberg}
\email{hirshb@tauex.tau.ac.il}
\affiliation{School of Chemistry, Tel Aviv University, Tel Aviv 6997801, Israel.}
\affiliation{Center for Computational Molecular and Materials Science, Tel Aviv University, Tel Aviv 6997801, Israel.}

\begin{abstract}
We propose an improved Path Integral Monte Carlo (PIMC) algorithm called Harmonic PIMC (H-PIMC) and its generalization, Mixed PIMC (M-PIMC). PIMC is a powerful tool for studying quantum condensed phases. However, it often suffers from a low acceptance ratio for solids and dense confined liquids. We develop two sampling schemes especially suited for such problems by dividing the potential into its harmonic and anharmonic contributions. In H-PIMC, we generate the imaginary time paths for the harmonic part of the potential exactly and accept or reject it based on the anharmonic part. In M-PIMC, we restrict the harmonic sampling to the vicinity of local minimum and use standard PIMC otherwise, to optimize efficiency.
We benchmark H-PIMC on systems with increasing anharmonicity, improving the acceptance ratio and lowering the auto-correlation time. For weakly to moderately anharmonic systems, at $\beta \hbar \omega=16$, H-PIMC improves the acceptance ratio by a factor of 6--16 and reduces the autocorrelation time by a factor of 7--30. We also find that the method requires a smaller number of imaginary time slices for convergence, which leads to another two- to four-fold acceleration. For strongly anharmonic systems, M-PIMC converges with a similar number of imaginary time slices as standard PIMC, but allows the optimization of the auto-correlation time. We extend M-PIMC to periodic systems and apply it to a sinusoidal potential. Finally, we combine H- and M-PIMC with the worm algorithm, allowing us to obtain similar efficiency gains for systems of indistinguishable particles. 
\end{abstract}

\maketitle

%\tableofcontents

\section{Introduction}

Path integral Monte Carlo (PIMC) is a powerful tool for studying the quantum properties of many body systems at thermal equilibrium. It has been widely applied to superfluid helium \cite{pollock1987path, ceperley1995path,Dmowski2017}, quantum liquids and solids \cite{ceperley2001path, boninsegni2006superglass, boninsegni2006fate, Boninsegni:2007sc,Soyler:2009pr, rota2011path,Herdman:2014qm,Borda:2016ds}, confined and low dimensional superfluids \cite{DelMaestro:2011ll,Kulchytskyy:2013dh,Markic:2015bu,Nava:2022qo,Happacher:2013pd,Kim:2024at}, warm dense matter \cite{dornheim2016ab, dornheim2018ab, dornheim2020nonlinear}, and ultra-cold gases \cite{Pilati:2006es,Trotzky2010,cinti2014defect, dornheim2020path, pascual2021quasiparticle, ghosh2024path}, to name a few.

At finite temperature $T$, inverse temperature $\beta = \left(k_{\text{B}}T\right)^{-1}$, the path integral expression for the partition function of a single particle in one spatial dimension with Hamiltonian $\hat H = \frac{\hat p^2}{2m} + \hat V$ is
\begin{align}\label{eq: Z_FP}
    Z(\beta) &= \text{Tr}\{e^{-\beta \hat H}\} \notag \\ &= \underset{P\rightarrow\infty}{\text{lim}} \int \dd{x_{1}} \dots \dd{x_{P}}\prod_{i=1}^{P} \langle x_{i}|e^{-\tau \hat H_0}|x_{i+1}\rangle e^{-\tau V(x_{i})},
\end{align}
where $\hat H_0 = \frac{\hat p^2}{2m}$ is the free particle Hamiltonian, $P$ is the number of imaginary time slices and $\tau = \frac{\beta}{P}$. For any finite $P$, the decomposition on the RHS of Eq.~\eqref{eq: Z_FP} is accurate only up to corrections of $\mathrm{O}(\tau)$ and is referred to as the primitive approximation.  In practice, one can employ decompositions of the density matrix that reduce the error to $\mathrm{O(\tau^4)}$ at the cost of evaluating derivatives of the potential \cite{Jang:2001jv}.  The sequence of positions, $\{ x_1, x_2, \cdots , x_P\}$, represents an imaginary time path associated with a particle, also called a worldline. The generalization of Eq.~\eqref{eq: Z_FP} to $N$ particles and $d$ spatial dimensions is straightforward. 

In PIMC, trial imaginary time paths are sampled from the free particle density matrix, $\rho_0(x_i,x_{i+1};\tau) \equiv \langle x_{i}|e^{-\tau \hat H_0}|x_{i+1}\rangle $, exactly due to its Gaussian form, accepting or rejecting them based on the change in potential between two paths. The expectation value of an observable $\hat A$ can then be obtained by averaging its corresponding estimator over the sampled paths. In the case that $\hat A$ is diagonal in the position representation, the corresponding estimator is $A_P(x_1, \cdots, x_P) = \frac{1}{P}\sum_{i=1}^P A(x_i)$. For momentum-dependent observables, the estimator is derived from the partition function (Eq.~\eqref{eq: Z_FP}) using thermodynamic relations, e.g., $\langle E \rangle = -\frac{\partial}{\partial \beta}  \text{ln}Z(\beta)$.

Algorithms to exactly sample free particles include bisection \cite{ceperley1995path} and staging \cite{sprik1985staging} which both take advantage of the result that the convolution of a Gaussian is also Gaussian and can thus be sampled exactly via Box-Muller \cite{Box:1958oz}.  PIMC was later extended via the worm algorithm~\cite{boninsegni2006worm1, boninsegni2006worm2}, allowing for simulations in the grand canonical ensemble and as well as efficient sampling of the worldline exchanges necessary to account for the indistinguishability of quantum particles. Despite these advancements, PIMC can suffer from a low acceptance probability of the proposed paths for dense systems at low temperatures \cite{boninsegni2006worm1,Zillich:2010hq,Mielke:2016vi,Yilmaz:2020ld} which can limit the efficiency of PIMC, and requires fine-tuning of the number of imaginary time slices that are modified in a proposed new configuration.  This can become problematic when PIMC is applied to quantum solids and dense confined liquids \cite{Kulchytskyy:2013dh,DelMaestro22,Rosenow:2024fo}.

In this paper, we introduce a method for proposing trial paths to improve the convergence of PIMC. For a given Hamiltonian, we separate the potential into its harmonic and anharmonic contributions. We propose paths by sampling the harmonic part exactly and accepting or rejecting the proposed path based on the anharmonic residual potential. 

This variant of PIMC, which we call Harmonic PIMC (H-PIMC), increases the acceptance ratio and significantly decreases the autocorrelation time between samples, especially at low temperatures. We demonstrate the efficiency of H-PIMC for weakly and moderately anharmonic systems, for which H-PIMC also reduces the number of imaginary time slices required for convergence. We then generalize H-PIMC by restricting the harmonic sampling to the vicinity of local minimum and combining it with standard PIMC elsewhere, which we call Mixed PIMC (M-PIMC). M-PIMC retains all the advantages of H-PIMC and, for strongly anharmonic systems, it also optimizes the autocorrelation time. We also extend M-PIMC to periodic systems and combine it with the worm algorithm to allow for the study of many indistinguishable particles, finding similar gains in efficiency. 

Several previous papers used some form of harmonic guidance to improve sampling. A harmonic reference system was shown to improve the convergence of observables with respect to the bead number \cite{Friesner:1984mo, Chao:1997gu}. The harmonically guided whole-path importance sampling of Mielke and Truhlar \cite{Mielke:2016vi} rejects almost all non-relevant paths at low temperatures without the expensive potential computation. The harmonic-phase approximation Monte Carlo approach of Robertson and Habershon \cite{Robertson:2017hl} improves the calculation of imaginary time-correlation function for anharmonic potential. More recently, Moustafa and Schultz \cite{Moustafa:2024kq} showed the advantage of updating the staging coordinates based on a harmonic reference potential in both PIMC and path integral molecular dynamics. However, surprisingly, the method was not incorporated in modern PIMC simulations using the worm algorithm, nor was its efficiency systematically analyzed going from model to periodic systems, as a function of anharmonicity, and including exchange statistics. The possibility of combining harmonic sampling near local minima with standard PIMC in other regions, as we propose in M-PIMC, have not been explored. We systematically address these aspects in this paper.

The paper is structured as follows. Section~\ref{sec: hpimc} provides the theory of H-PIMC. M-PIMC and its extension to periodic systems are discussed in Section~\ref{sec: mpimc} and Section~\ref{sec: mpimcpbc}, respectively. Section~\ref{sec: mpimc_worm} combines H-PIMC and M-PIMC with the worm algorithm. We benchmark H-PIMC on harmonic and anharmonic potentials with and without the worm algorithm in Section~\ref{sec: HPIMC_res}. Section~\ref{sec: AHOresultsMPIMC} discusses the application of M-PIMC to the strongly anharmonic system. The application of M-PIMC to a periodic system is discussed in Section~\ref{sec: MPIMC_PBC_results}. Finally, we conclude the paper in Section~\ref{sec: summary}. 

\section{Theory}
\subsection{Harmonic PIMC}\label{sec: hpimc}
For a given external 1-body potential, it is possible to separate the harmonic and anharmonic contributions, $V(x) = V_{\text{ho}}(x) + V_{\text{anh}}(x)$, where $V_{\text{ho}}(x)= \frac{1}{2} m \omega^2 x^2$ and $V_{\text{anh}}(x)$ are the harmonic and anharmonic parts, respectively, and $\omega = \sqrt{\frac{1}{m} \dv[2]{V(x)}{x}}\big\rvert_{x = x_{\text{min}}}$ is the curvature near the potential minimum ($x_{\text{min}}$).
We can then rewrite the Hamiltonian as $\hat H = \hat H_0 + \hat V = \hat H_{\text{ho}} + \hat V_{\text{anh}}$, where $\hat H_{\text{ho}} = \hat H_0 + \hat V_{\text{ho}}$. 
As a result, Eq.~\ref{eq: Z_FP} becomes,
\begin{equation}\label{eq: Z_HO}
Z(\beta) = \underset{P\rightarrow\infty}{\text{lim}} \int \dd{x_{1}} \dots \dd{x_{P}}\prod_{i=1}^{P} \langle x_{i}|e^{-\tau \hat H_{\text{ho}}}|x_{i+1}\rangle e^{-\tau V_{\text{anh}}(x_{i})},
\end{equation}
where the harmonic oscillator density matrix \cite{FeynmanHibbs2010,Shao:2016eu, Barragan:2018px}, 
\begin{align}
    &\rho_{\text{ho}}(x_i, x_{i+1};\tau) \equiv \langle x_{i}|e^{-\tau \hat H_{\text{ho}}}|x_{i+1}\rangle \nonumber \\
&= \sqrt{\frac{m\omega}{2\pi\hbar\sinh(\tau\hbar\omega)}}
\mathrm{e}^{ -\frac{m\omega}{2\hbar\sinh(\tau\hbar\omega)}
    \bqty{(x_i^2 + x_{i+1}^2)\cosh(\tau\hbar\omega)
- 2x_i x_{i+1}}}
\end{align}
can be sampled directly. Note that we obtained Eq.~\eqref{eq: Z_HO} using a symmetric Trotter splitting $e^{-\tau \hat H} \approx e^{-\frac{\tau}{2} \hat V_{\text{anh}}}e^{-\tau \hat H_{\text{ho}}}e^{-\frac{\tau}{2} \hat V_{\text{anh}}}$, which we refer to as ``harmonic Trotter splitting''. In contrast, we used the splitting $e^{-\tau \hat H} \approx e^{-\frac{\tau}{2} \hat V} e^{-\tau \hat H_0}e^{-\frac{\tau}{2} \hat V}$ to obtain Eq.~\eqref{eq: Z_FP}, which we call ``free particle Trotter splitting,'' ignoring errors of $\mathrm{O}(\tau^2)$.

The form of the partition function in Eq.~\eqref{eq: Z_HO} provides an alternative sampling method to standard PIMC using harmonic oscillator paths as trial paths. We refer to this approach as harmonic PIMC (H-PIMC) which involves proposing imaginary time paths based on exactly sampling the harmonic part of the Hamiltonian, and performing a Metropolis accept or reject based on the anharmonic potential. The acceptance probability in H-PIMC for a Monte Carlo update of the position of a single bead $\{x_1,..., x,..., x_P\} \rightarrow \{x_1,..., y,..., x_P\}$ is
\begin{equation}\label{eq: accprob_hpimc}
    A(y|x) = \text{min} \Bigg[1, \frac{e^{-\tau V_{\text{anh}}(y)}}{e^{-\tau V_{\text{anh}}(x)}} \Bigg]. 
\end{equation} 
In practice, new positions are proposed for all beads at every Monte Carlo step.
We anticipate this sampling method to be most efficient at low temperatures, when the system primarily explores regions near a single potential minimum, and fluctuations around it contribute the most to the partition function. 

The partition functions in Eq.~\eqref{eq: Z_FP} and Eq.~\eqref{eq: Z_HO} are equivalent only in the limit $P\to\infty$. At finite $P$, they lead to different estimators for the total energy. It can be shown that the estimators corresponding to position-dependent diagonal observables in H-PIMC are the same as in standard PIMC. For momentum-dependent observables, we derive new estimators using thermodynamic relations. For example, the total energy estimator becomes,
\begin{align}\label{eq:Energy_hpimc}
    &\epsilon_{P}^{\text{(H-PIMC)}}(x_1, \cdots, x_P) = \sum_{i=1}^{P} \Bigg[ \frac{\hbar\omega}{2P\tanh(\tau\hbar\omega)} \notag \\
    &-\frac{m\omega}{2\hbar}\frac{\hbar\omega/P}{\sinh^{2}(\tau\hbar\omega)}\left(x_{i}^{2}+x_{i+1}^{2}\right) \notag \\ 
    &+\frac{m\omega}{2\hbar}\frac{\hbar\omega/P}{\tanh(\tau\hbar\omega)\sinh(\tau\hbar\omega)}2x_{i}x_{i+1} 
    +\frac{1}{P}V_{\text{anh}}(x_{i}) \Bigg].
\end{align}

As we show in Section~\ref{sec:results}, H-PIMC is very efficient for weakly to moderately anharmonic systems. However, H-PIMC loses its advantage with increasing anharmonicity. For strong anharmonicities, we combine H-PIMC near the potential minimum with standard PIMC in other regions, as we describe next.

%---------------------------------------------------------------------------------
\subsection{Combining PIMC and H-PIMC}\label{sec: mpimc}
\begin{figure}[ht]
\centering
\includegraphics[width=0.475\textwidth]{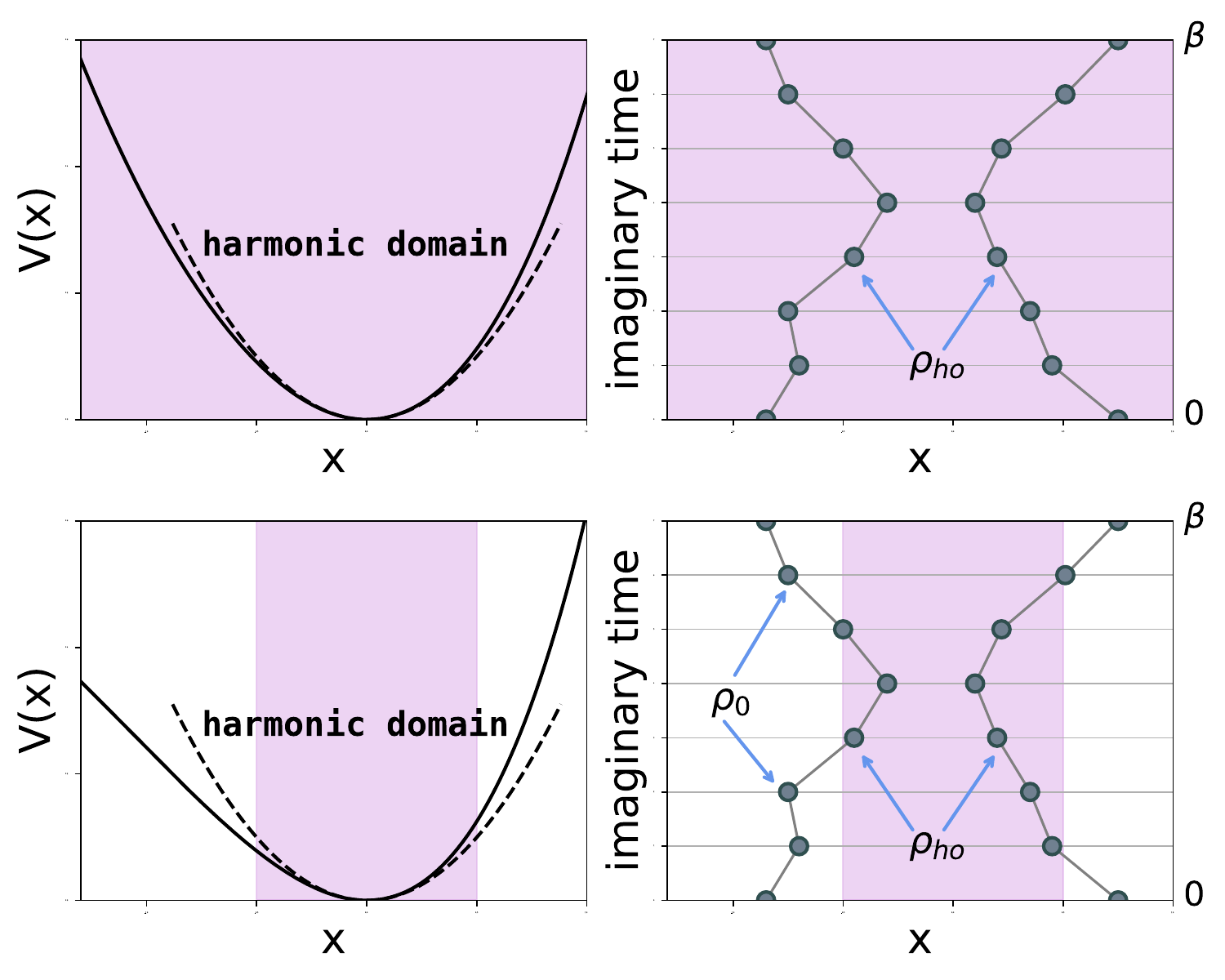}
\caption{A schematic of the H-PIMC (upper panel) and M-PIMC (lower panel) algorithms. The full potential and its harmonic approximation are shown as solid and dashed lines, respectively (left). The shaded region indicates the harmonic domain. In H-PIMC, all the bead locations are proposed using a harmonic update. In M-PIMC, for a given system configuration, some beads fall within the harmonic domain (shaded region). For these, trial positions are sampled using the harmonic update, as in H-PIMC. For the remaining beads, trial positions are sampled using the free particle density matrix, as in PIMC.}
\label{fig:mpimc_cartoon}
\end{figure}

With increasing anharmonicity, the anharmonic part of the potential $V_{\text{anh}}(x)$ begins to dominate over the harmonic contribution $V_{\text{ho}}(x)$. As a result, the fluctuations of the system are harmonic only in a very close proximity to the potential minimum, which we call the harmonic domain, and the harmonic paths are not good trial paths across the entire configuration space.  As a result, we propose the following mixed trail path approach: We define a harmonic domain close to the potential minimum (see Figure~\ref{fig:mpimc_cartoon}), and propose harmonic paths only for the beads that lie within it, as in H-PIMC. Otherwise, for the beads outside the harmonic domain, the trial positions are proposed using the free particle density matrix, as in standard PIMC. Note that we can use these ``mixed trial paths'' together with either the Trotter splitting of Eq.~\eqref{eq: Z_FP} or \eqref{eq: Z_HO}. Throughout this work, unless noted otherwise, we use the splitting of Eq.~\eqref{eq: Z_HO} with the mixed trial path proposal, as it offers benefits that will become clear in subsequent sections. This approach generalizes the H-PIMC algorithm, allowing for broader applicability even in the absence of a strongly harmonic potential. We call this method mixed PIMC (M-PIMC). Figure~\ref{fig:mpimc_cartoon} shows the M-PIMC algorithm schematically in comparison to H-PIMC.

Consider a simple Monte Carlo update $\{x_i, x, x_f\} \rightarrow \{x_i, y, x_f\}$ with fixed ends $x_{i}$ and $x_{f}$, separated by imaginary time 2$\tau$. Since two different types of moves are possible, the forward move ($x \rightarrow y$) and the reverse move ($y \rightarrow x$) might not be of the same kind, e.g., if $x$ is within the harmonic domain and $y$ is outside it. In M-PIMC, we fix the choice of the trial probability, $T(y|x) \equiv T(x\rightarrow y; x_i, x_f)$, based on the bead location $x$. If $x$ is within the harmonic domain, then the trial update $x \rightarrow y$ is sampled using the harmonic action:
\begin{equation}
    T(y|x) := T_{\text{ho}}(y|x) = \frac{1}{\sqrt{2 \pi \sigma_{\text{ho}}^2}} e^{-\frac{(y - \bar y_{\text{ho}})^2}{2\sigma_{\text{ho}}^2}},
\end{equation}
where $\bar y_{\text{ho}} = \frac{\Gamma_2}{\Gamma_1}$, $\sigma_{\text{ho}}^2 = \frac{\hbar}{m \omega \Gamma_1}$, $\Gamma_1 = 2 \coth(\tau \hbar \omega)$ and $\Gamma_2 = \frac{x_i + x_f}{\sinh(\tau \hbar \omega)}$. However, if $x$ is outside the harmonic domain, then the trial update $x \rightarrow y$ is sampled using free particle action, i.e.,
\begin{equation}
    T(y|x) := T_{0}(y|x) = \frac{1}{\sqrt{2 \pi \sigma_{0}^2}} e^{-\frac{(y - \bar y_{0})^2}{2\sigma_{0}^2}},
\end{equation}
where $\bar y_{0} = \frac{x_i + x_f}{2}$, $\sigma_{0}^2 = \lambda \tau$ and $\lambda = \frac{\hbar^2}{2m}$. Given $T(y|x)$, we propose a trial move, and $T(x|y)$ is similarly determined from the position of the proposed $y$.

As a result, the acceptance probability for the harmonic Trotter splitting (as in Eq.~\eqref{eq: Z_HO}) is
\begin{equation}\label{eq: accprob_mpimc_HS}
    A(y |x) = \text{min} \Bigg[1, \frac{T(x|y)/T_{\text{ho}}(x|y)}{T(y|x)/T_{\text{ho}}(y|x)} \frac{e^{-\tau V_{\text{anh}}(y)}}{e^{-\tau V_{\text{anh}}(x)}} \Bigg].
\end{equation}
Note that when $x$ and $y$ are both in the harmonic domain, or if the harmonic domain extends throughout the whole configuration space, we recover Eq.~\eqref{eq: accprob_hpimc}.
On the other hand, if $x$ and $y$ are both are outside the harmonic domain, the acceptance probability becomes equivalent to standard PIMC in the $\tau \to 0$ limit (see Appendix~\ref{app:MPIMC_Acceptance_Ratio}).
The acceptance probability for the free particle Trotter splitting (as in Eq.~\ref{eq: Z_FP}) becomes
\begin{equation}\label{eq: accprob_mpimc_FPS}
    A(y |x) = \text{min} \Bigg[1, \frac{T(x|y)/T_{0}(x|y)}{T(y|x)/T_{0}(y|x)} \frac{e^{-\tau V(y)}}{e^{-\tau V(x)}} \Bigg].
\end{equation}
In this case, when both $x$ and $y$ are outside the harmonic domain, or the harmonic domain shrinks to zero, we recover the traditional acceptance probability for standard PIMC. Details of the derivation of these expressions are given in Appendix~\ref{app:MPIMC_Detailed_Balance}.

The generalization of Eq.~\eqref{eq: accprob_mpimc_HS} (or Eq.~\eqref{eq: accprob_mpimc_FPS}) to a full path update $\{x_1, x_2, \cdots, x_{P-1}, x_P\} \rightarrow \{x_1, y_2, \cdots, y_{P-1}, x_P\}$ is straightforward and used in all simulation reported in the paper.

% ---------------------------------------------------------------------
\subsection{M-PIMC for periodic systems}\label{sec: mpimcpbc}
To address the case of multiple minima, we consider a particle in a one-dimensional periodic potential, $V(x)$, with periodicity $L$. The partition function is given by \cite{krauth2006statistical, pollock1987path, spada2022path},
\begin{align}\label{eq:Z_pbc}
    & Z^{\text{(PBC)}}(\beta) = \underset{P\rightarrow\infty}{\text{lim}} \prod_{i=1}^{P} \int \dd{x_i} \langle x_i | \mathrm{e}^{-\tau \hat H_0} | x_{i+1} \rangle^{\text{(PBC)}} e^{-\tau V(x_i)} \notag \\
    &= \underset{P \rightarrow \infty}{\text{lim}} \prod_{i=1}^{P} \int \dd{x_i} \sum_{w_i = - \infty}^{\infty} \langle x_i | \mathrm{e}^{-\tau \hat H_0} | x_{i+1} + w_{i+1} L \rangle e^{-\tau V(x_i)},
\end{align}
where $\langle x_i | e^{-\tau \hat H_0} | x_{i+1} \rangle^{\text{(PBC)}}$ is the density matrix of a free particle inside a periodic box of length $L$ at inverse temperature $\tau$. The bead positions $\{x_1, x_2, \cdots , x_P \}$ lie within the fundamental box, and $w_{i+1}$ is an integer called the local winding number associated with the bond between the $i$-th and $(i+1)$-th beads. The total winding number associated with the full path is $W \equiv \frac{1}{L} \int_0^{\hbar \beta} \dd{\tau} \dv{x(\tau)}{\tau} = \sum_{i=1}^{P} w_i$ represents the number of times the path \emph{wraps} around the periodic boundary conditions. In PIMC, we first sample the total winding number, $W$, then propose a free particle trial path that satisfies the $W$-constraint, e.g., using tower sampling~\cite{krauth2006statistical}, and accept or reject it based on change in potential between the two paths.

Harmonic paths typically fluctuate around the minima, therefore using H-PIMC and M-PIMC will not sample non-zero winding numbers efficiently. Therefore, we propose the mixed trial path proposal, as in M-PIMC, for paths with $W = 0$ and the free particle trial path for the worldlines with $W \neq 0$, as in standard PIMC. We call this algorithm M-PIMC-PBC. Note that in M-PIMC-PBC, we use the free particle Trotter splitting to derive Eq.~\eqref{eq:Z_pbc}.

\subsection{The worm algorithm}\label{sec: mpimc_worm}
Until this point we have only considered a single particle $N=1$.  To extend this work to $N>1$, we combine M-PIMC with the worm algorithm \cite{boninsegni2006worm1, boninsegni2006worm2} that allows efficient simulation of thousands of identical particles. The worm algorithm extends the PIMC configuration space by considering a single partial worldline (the worm) that does not wrap around the $\beta$-cylinder. Configurations with a worm are referred to as off-diagonal and contribute to the single particle Green function. Configurations without the worm contribute to the partition function of the whole system. Including a worm allows to extend PIMC to the grand canonical ensemble, but since we are only interested in canonical measurements we consider a subset of worm configurations where the number of particles in the system remains fixed. Formally, we can generalize the single particle partition function in Eq.~\eqref{eq: Z_HO} for many particles as
\begin{equation}\label{eq: Z_FP_wSymmetry}
    Z_N = \frac{1}{N!}\underset{P\rightarrow\infty}{\text{lim}}\sum_{\sigma} \int \prod_{i=1}^{P} \dd{r_{i}} \langle r_{i}|e^{-\tau \hat H_{\text{ho}}}|r_{i+1}\rangle e^{-\tau V_{\text{anh}}(r_{i})}
\end{equation}
where $r_i = (x_{\alpha_1,i} .... x_{\alpha_N,i})$ and $\sum_{\sigma}$ refers to the sum over all possible permutations of the $x$'s. Here, $\alpha_i$ are the N particle indices. The off-diagonal sector is characterized by a single extra bead $x_H$ at some time slice $m$ for some particle index $\alpha_H$. Since we are in the canonical ensemble, the number of links must be preserved i.e. there is a link between $x_{\alpha_H,m-1}$ and $x_H$ and one between $x_{\alpha_H,m}$ and $x_{\alpha_H,m+1}$. Analogously we can define an off-diagonal partition function,
\begin{align}
Z' &= \frac{1}{N!}\underset{P\rightarrow\infty}{\text{lim}}\sum_{\sigma}\sum_{\alpha_H = 1}^N\int_V dx_H \int \prod_{i=1}^{P} \dd{r_{i}} \langle r_{i}|e^{-\tau \hat H_{\text{ho}}}|r_{i+1}\rangle \notag \\
&\times e^{-\tau V_{\text{anh}}(r_{i})},
\end{align}
and total partition function is thus $Z \equiv  Z_N + CZ'$. Here, $C$, the worm constant is a hyperparameter that controls the ratio between the time the code spends in the diagonal and off-diagonal sectors. We find that tuning at runtime to obtain a ratio $\sim 75 \% $ is optimal.

To incorporate harmonic sampling into the worm algorithm, which can then be combined seamlessly with H-PIMC, we need to introduce and modify three worm updates: Open, Close and Swap \cite{boninsegni2006worm2}. 

\subsubsection{Open and Close} 

These are complimentary updates which form a detailed balance pair. Open works by breaking the link between two beads and sampling the position of a single new bead (called the head) which lies on the same time slice. The old bead is referred to as the tail. (Historically, they are known in the literature as Ira and Masha). The acceptance ratio of the open move i.e. to go from a diagonal configuration $x$ in $Z$ to an off-diagonal one $y$ in $Z^\prime$:  
\begin{equation}
    A(y|x) = \text{min}\left[1,\frac{N P C}{\rho_{\text{ho}}(x_{\alpha_H,m-1},x_H,\tau)}e^{\Delta V_{\text{anh}}}\right]
\end{equation} 
Here, $NP$ corresponds to the number of beads in $Z_N$, the diagonal configuration and $\Delta V_{\text{anh}}$ is the total change in anharmonic energy $V_{\text{anh}}(y) - V_{\text{anh}}(x)$.

\subsubsection{Swap} 
The Swap move is the key addition which allows us to sample permutations of identical particles efficiently. It consists of linking the head to a different worldline and creating a new head bead in the process which lies on a different world line. This allows us to sample configurations in different global winding sectors with only spatially local updates and avoids large potential barriers resulting from the presence of hard-core interactions.   This update can happen only in the off-diagonal configuration $Z^\prime$.
\begin{equation}
        A(y|x) = \text{min}\left[1,\frac{\Sigma_W}{\Sigma_K}e^{\Delta V_{\text{anh}}}\right]
\end{equation}
Here $\Sigma_W = \sum_{i=1}^{N} \rho_{\text{ho}}(x_{H},x_{\alpha_i,m+j},j\tau)$ where the worm head is at $m$ and $j$ is the number of beads involved in the update. $j$ is another hyperparamter that can be optimized to improve the acceptance ratio of the Swap move.  In practice, for non-interacting particles, we choose $j=1$. Similarly, $\Sigma_K = \sum_{i=1}^{N} \rho_{\text{ho}}(x_{\alpha_k,m},x_{\alpha_i,m+j},j\tau)$ where $k$ is the chosen bead for swapping.

Having defined multiple variants of PIMC with improved sampling to take advantage of spatially local quasi-harmonic confinement, we now benchmark our algorithm on a number of systems with $N=1,2$. 

%=========================================================
\section{Results}\label{sec:results}
%=========================================================

To obtain a baseline for comparison, we first apply H-PIMC to harmonic, weakly anharmonic and strongly anharmonic model potentials. Then, we show that improved sampling can be obtained for strongly anharmonic systems using M-PIMC. We also apply M-PIMC-PBC to a sinusoidal trap.
To evaluate the efficiency of our proposed algorithm, we compute the staging acceptance ratio, defined as the ratio of the total number of accepted staging moves and the total number of attempted staging moves. We also compute the average total energy and estimate the energy autocorrelation time. We can define the integrated autocorrelation time for an observable $A$ as,
\begin{equation}
    \tau_{\text{int}}^{(A)} = 1 + 2\sum_{t=1}^{M} \frac{C^{(A)}(t)}{C^{(A)}(0)}
\end{equation}
where $M$ is the number of Monte Carlo steps and $C^{(A)}(t)$ is defined as
\begin{equation*}
C^{(A)}(t) = \frac{1}{M - t} \sum_{i=1}^{M - t} 
\bigl( A(i) - \langle A \rangle \bigr) 
\bigl( A(i + t) - \langle A \rangle \bigr).
\end{equation*}
We use the emcee Python library \cite{Foreman-Mackey_2013} to estimate the integrated autocorrelation time.

\begin{figure}[ht]
\centering
\includegraphics[width=0.35\textwidth]{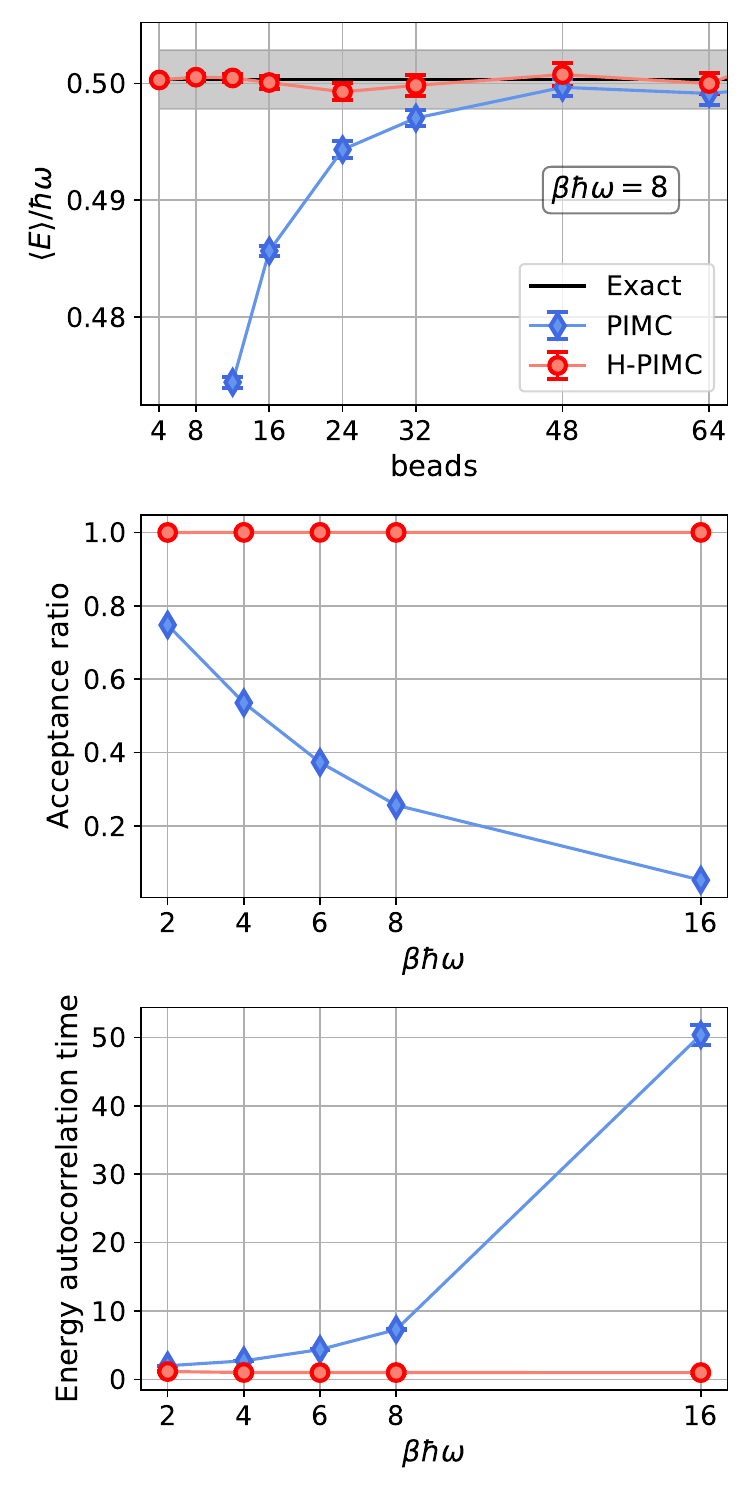}
\caption{Comparison between PIMC and H-PIMC for harmonic potential. The energy convergence with the number of beads is shown for both PIMC and H-PIMC at $\beta \hbar \omega = 8$. The shaded region indicates a 0.5$\%$ deviation from the exact energy. The acceptance ratio and the energy autocorrelation time data are obtained using the number of beads required for energy convergence at each temperature. For example, at $\beta \hbar \omega =8$, the number of beads used in PIMC and H-PIMC are 48 and 8, respectively. Details of the parameters are given in Appendix~\ref{app:Params}.}
\label{fig:hpimc_sho}
\end{figure}
\subsection{Application of H-PIMC}\label{sec: HPIMC_res}

\subsubsection{For a harmonic potential}\label{sec: HOresultHPIMC}
Since H-PIMC samples the density matrix of the quantum harmonic oscillator exactly, it is an exact method for a harmonic trap without interactions: 
\begin{equation}
\hat{V} = \frac{1}{2} m \omega^2 \hat{x}^2 
\end{equation}
Figure~\ref{fig:hpimc_sho} shows the H-PIMC results and its comparison with standard PIMC for a particle inside a harmonic trap at various temperatures ($\beta \hbar \omega$). The results show the energy convergence with the number of beads for both PIMC and H-PIMC at $\beta \hbar \omega = 8$. H-PIMC being an exact method for this system, the average energy is independent of the number of beads, whereas PIMC requires at least 48 beads for energy convergence within statistical uncertainty. Also, since H-PIMC generates harmonic fluctuations, the acceptance rate is $100\%$ for all temperatures as seen in the middle panel. In contrast, the acceptance ratio for PIMC decreases with temperature. H-PIMC also significantly improves energy autocorrelation time. The improvements are more prominent at lower temperatures. For example, at $\beta \hbar \omega = 8$, and $16$, the improvement in energy autocorrelation time is by a factor of 8 and 50, respectively.

\subsubsection{For an anharmonic potential}\label{sec: AHOresultsHPIMC}
Next we apply the H-PIMC algorithm to an anharmonic system with increasing anharmonicity. The anharmonic potential is of the form
\begin{equation}
   \hat V = \frac{1}{2} m \omega^2 \hat{x}^2 \left(1 + c_3 \hat{x} + c_4 \hat{x}^2 \right),
\end{equation}
where $c_3$, $c_4$ determine the level of anharmonicity. Here, we consider three different anharmonicity regimes: weekly, moderately and strongly anharmonic regime (Details of the parameters are given in Appendix~\ref{app:Params}).

Figure~\ref{fig:hpimc_res_bhw16} shows the H-PIMC results with increasing anharmonicity and its comparison with standard PIMC. For weakly anharmonic systems (upper panel), H-PIMC is significantly more efficient than PIMC. The average energy converges much faster with the number of beads for H-PIMC. For example, at $\beta \hbar \omega = 8$, H-PIMC converges with around 20 beads, whereas PIMC requires around 48 beads for energy convergence.  H-PIMC also significantly increases the acceptance ratio and reduces the energy autocorrelation time. Specifically, at low temperatures, $\beta \hbar \omega = 8$ and $16$, the improvement in energy autocorrelation time is by a factor of 5 and 30, respectively. Note that the proposal of the harmonic trial paths is responsible for the improvements in the acceptance ratio and the energy autocorrelation time, but the faster convergence of the energy with the number of beads is due to the energy estimator resulting from the harmonic Trotter splitting.

For moderate anharmonicity (middle row), H-PIMC again shows better efficiency than PIMC. H-PIMC improves the acceptance ratio and the energy autocorrelation time particularly at low temperatures $\beta \hbar \omega = 8, 16$. The average energy also converges faster with the number of beads. Overall, for weakly to moderately anharmonic systems, at the lowest temperature $\beta \hbar \omega=16$, H-PIMC improves the acceptance ratio by a factor of 6--16 and reduces the autocorrelation time by a factor of 7--30.

However, with increasing anharmonicity, H-PIMC loses its advantage gradually. For strong anharmonicity (lower panel), the energy convergence with the number of beads is similar for PIMC and H-PIMC and the improvement in acceptance ratios is only marginal. Furthermore, H-PIMC slightly increases the energy autocorrelation time, except at the lowest temperature considered here, $\beta \hbar \omega = 16$. 
\begin{figure*}[t]
\centering
\includegraphics[width=1.0\textwidth]{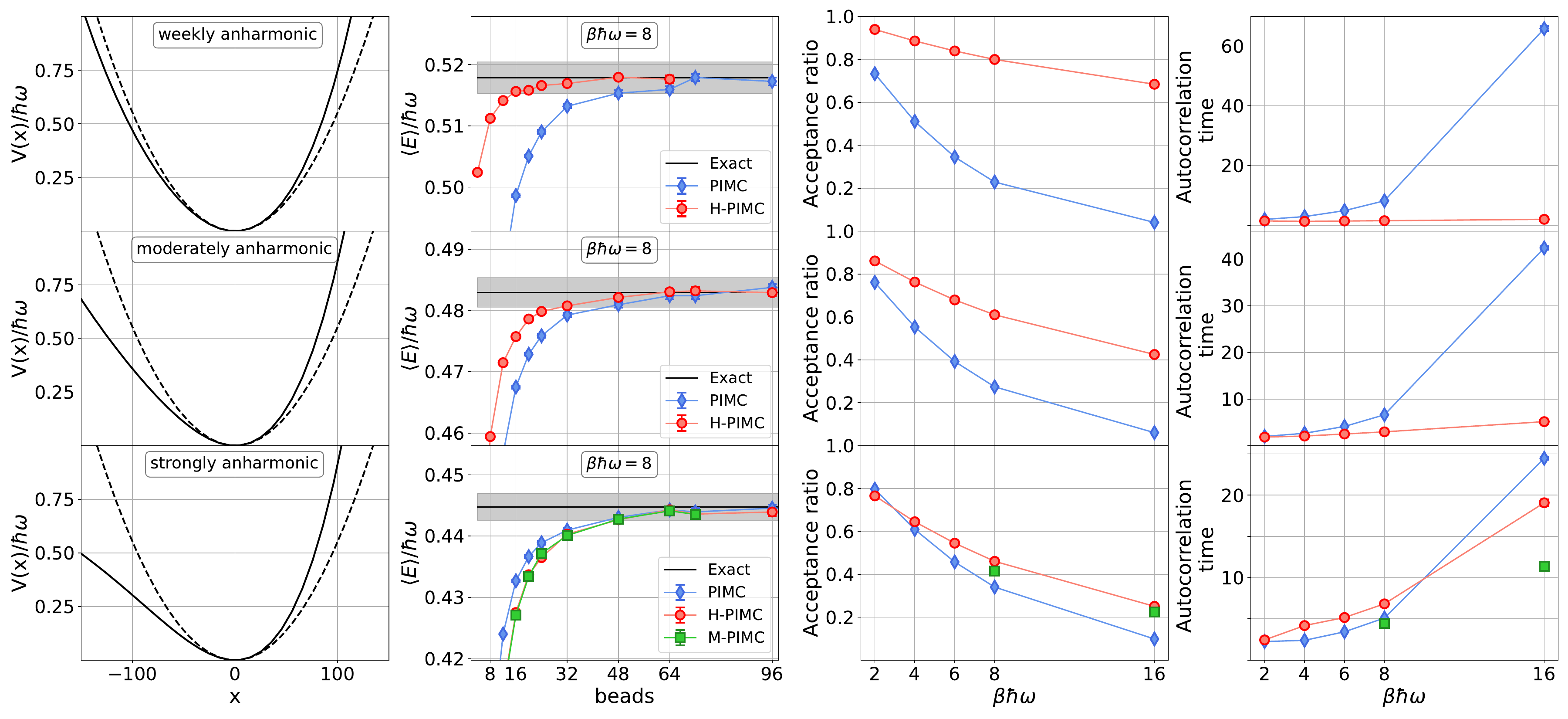}
\caption{A comparison between PIMC and H-PIMC results is shown for a weakly anharmonic system (upper panel), a moderately anharmonic system (middle panel), and a strongly anharmonic system (lower panel). The anharmonic potential and its harmonic approximation are shown in solid and dashed lines, respectively (first column). The average energy convergence is shown in the second column. The shaded region indicates a $0.5 \%$ deviation from the exact energy. The acceptance ratio (third column) and the energy autocorrelation time data (fourth column) are obtained using the number of beads required for energy convergence at each temperature. For the strongly anharmonic case (lower panel), the M-PIMC results for the optimized harmonic domain are shown as green squares.}
\label{fig:hpimc_res_bhw16}
\end{figure*}

\subsubsection{For $N=2$ indistinguishable particles}
\begin{figure*}[t]
\centering
\includegraphics[width=1.0\textwidth]{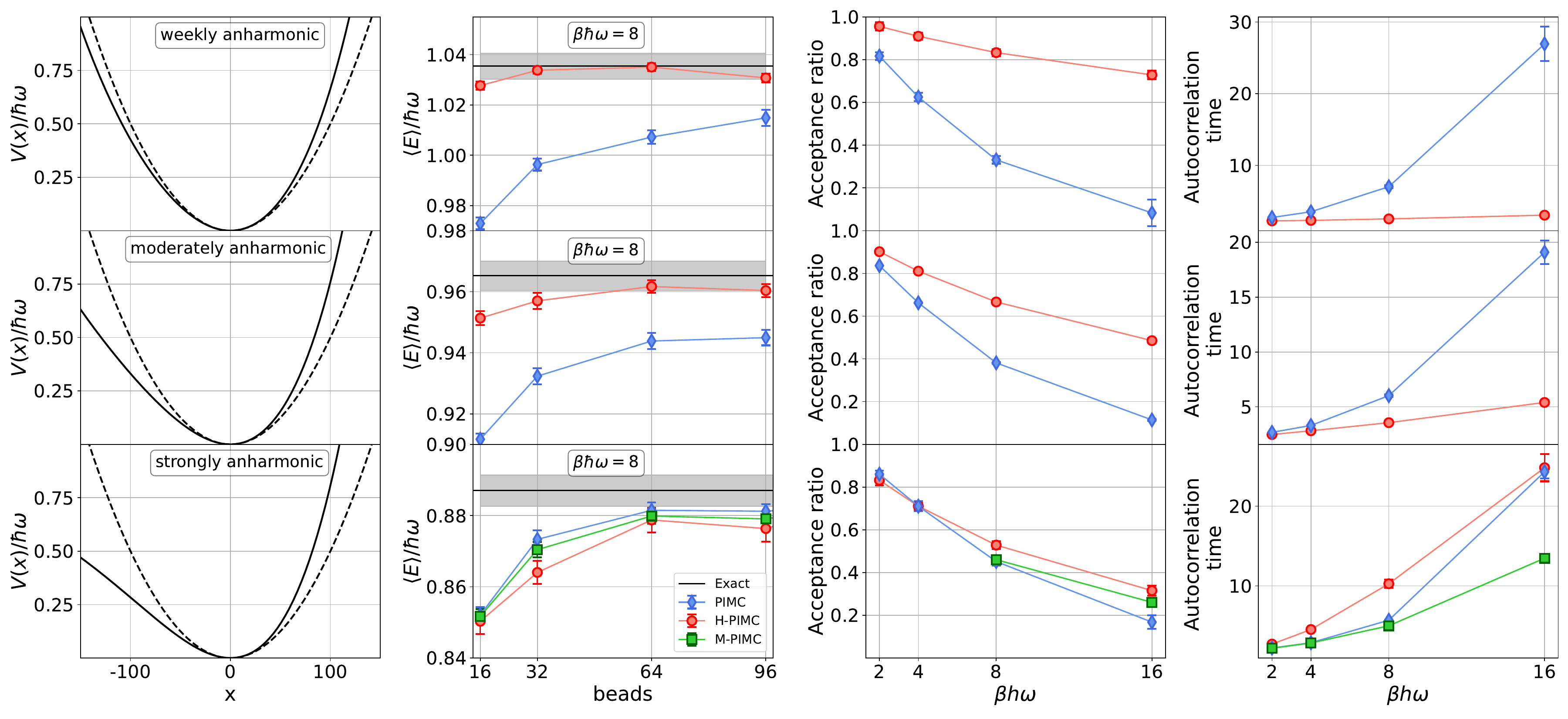}
\caption{A comparison between PIMC and H-PIMC results for N = 2 indistinguishable particles is shown for a weakly anharmonic system (upper panel), a moderately anharmonic system (middle panel), and a strongly anharmonic system (lower panel). The anharmonic potential and its harmonic approximation are shown in solid and dashed lines, respectively (first column). The average energy convergence is shown in the second column. The shaded region indicates a $0.5 \%$ deviation from the exact energy. The acceptance ratio (third column) and the energy autocorrelation time data (fourth column) are obtained using the number of beads required for energy convergence at each temperature.}
\label{fig:hpimc_res_N2}
\end{figure*}

H-PIMC provides similar improvements as in the single particle case also for indistinguishable particles. 
In Figure \ref{fig:hpimc_res_N2} we observe the same improvements in acceptance ratio, autocorrelation time and convergence with the number of beads for $N=2$ bosons trapped via weakly to moderately anharmonic one-dimensional potentials. These improvements gradually disappear for strongly anharmonic confinement, as before, for which we next apply M-PIMC.

\subsection{M-PIMC for strong anharmonicity}\label{sec: AHOresultsMPIMC}
\begin{figure}[ht]
\centering
\includegraphics[width=0.45\textwidth]{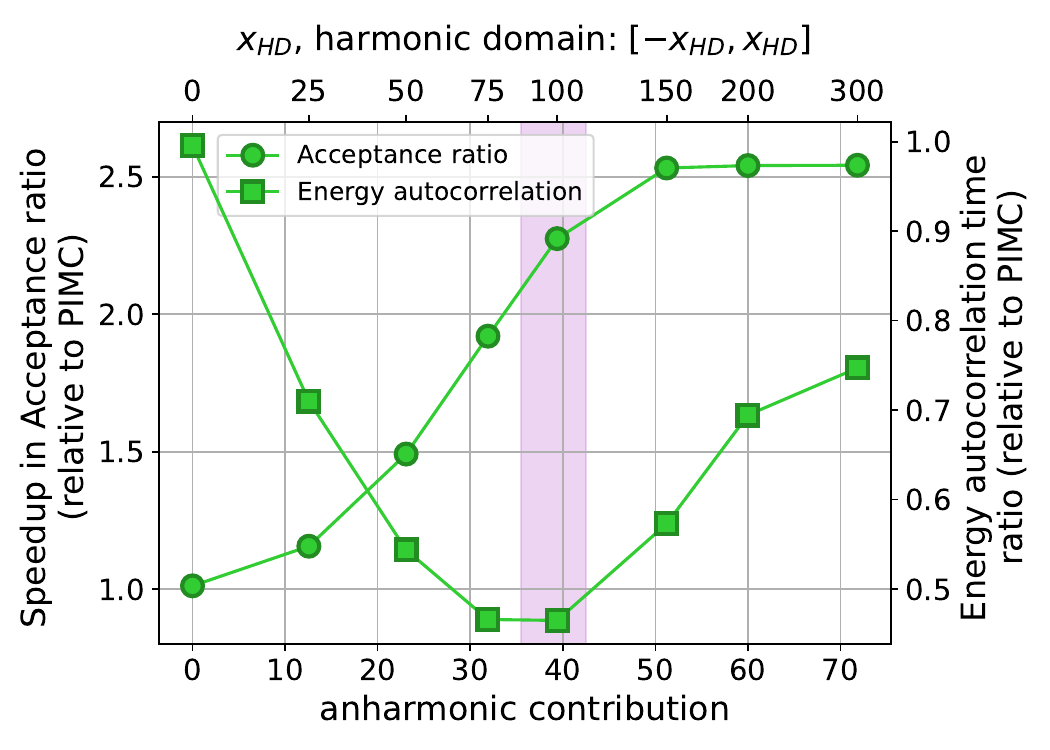}
\caption{M-PIMC results for strong anharmonicity at $\beta \hbar \omega = 16$. The M-PIMC calculations use 96 beads. The speedup in acceptance ratio is defined as $\frac{\text{M-PIMC value}}{\text{PIMC value}}$ and the energy autocorrelation time ratio is defined as $\frac{\text{M-PIMC value}}{\text{PIMC value}}$. The shaded area indicates the optimal harmonic domain.}
\label{fig:mpimc_res}
\end{figure}

We now apply M-PIMC to the strongly anharmonic system to address the inefficiency of H-PIMC for such systems. Figure~\ref{fig:mpimc_res} shows the improved acceptance ratio and autocorrelation time in M-PIMC for the strongly anharmonic case and $\beta \hbar \omega = 16$. The results are plotted as a function of the anharmonic contribution, which is defined for a given harmonic domain $[x_{\text{min}}-x_{\text{HD}},x_{\text{min}}+ x_{\text{HD}}]$, as $\frac{100 \times |V_{\text{anh}}(x_{\text{HD}})|}{|V(x_{\text{HD}})|}$. In this case, the anharmonic contribution increases monotonically with the size of the harmonic domain since regions further from the local minimum are included. Figure~\ref{fig:mpimc_res} shows that the acceptance ratio increases monotonically with the size of the harmonic domain, whereas the energy autocorrelation time shows nonmonotonic behavior. For a smaller harmonic domain, most of the bead updates are free particle updates, as in standard PIMC, with a relatively higher autocorrelation time. On the other extreme, for a very large harmonic domain, most of the bead updates are harmonic updates and the autocorrelation is again high, as in H-PIMC for strong anharmonicity. M-PIMC thus offers a route to optimize the energy autocorrelation time by tuning the size of the harmonic domain. In the optimal harmonic domain, we sample using the harmonic move only those beads which are in the vicinity of local minima, and, therefore, minimize the overall energy autocorrelation time. The results in Fig.~\ref{fig:mpimc_res} suggest that a harmonic domain with an anharmonic contribution of $35-45\%$ provides an optimal domain where the energy autocorrelation time is minimum, which also has a better acceptance ratio than PIMC. The improvement becomes more significant with decreasing temperature. The acceptance ratio and the energy autocorrelation time using M-PIMC and the optimal harmonic domain for more temperatures are shown in Figs.~\ref{fig:hpimc_res_bhw16}-\ref{fig:hpimc_res_N2} (green squares) for distinguishable and two indistinguishable particles, respectively, in a strongly anharmonic potential.
The above conclusions hold as we increase the number of particles (See Appendix~\ref{app: boson_speedup}). Thus, the optimized harmonic domain from single particle results are anticipated to work for many independent particles. 

\begin{figure}[ht]
\centering
\includegraphics[width=\columnwidth]{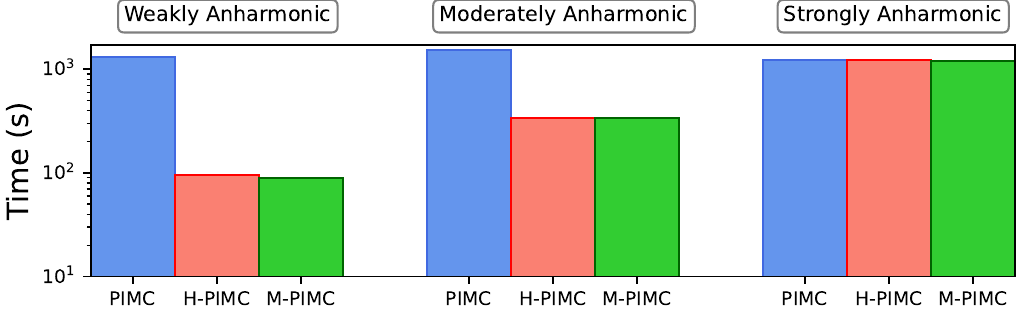}
\caption{Comparison of wall-time to solution between the three different methods for the three different cases illustrated in this paper. The time is taken to be the wall-time in seconds taken to achieve an error of 0.1\% for an individual simulation.}
\label{fig:wallclock}
\end{figure}

We quantify the speedup possible via the algorithms proposed here also in terms of code run-time (wall-time). In Fig. \ref{fig:wallclock}, we compare the time to solution between the three methods, H-PIMC, M-PIMC and standard PIMC for two bosonic particles in an anharmonic potential with three different strengths of anharmonicity as before. For weakly and moderately anharmonic case, we see improvements for using M-PIMC and H-PIMC over standard PIMC. For the strongly anharmonic case, despite M-PIMC reducing the autocorrelation times and improving the acceptance ratio over H-PIMC (see Fig.~\ref{fig:mpimc_res_2boson}), the wall-time to solution is similar. This is a result of increased logic in the code, which will be further optimized in the future.
We find that speedups of approximately an order of magnitude are possible for weakly and moderately anharmonic systems using H-PIMC.

\subsection{M-PIMC-PBC for sinusoidal potential}\label{sec: MPIMC_PBC_results}
We apply M-PIMC-PBC to a sinusoidal potential $V(x) = -V_0 \cos{(\frac{2\pi x}{L})}$ with varying values of $V_0$. For a chosen box length $L$, we consider three different barrier height regimes, $V_0$: low, intermediate and high (Details of the parameters are given in Appendix~\ref{app:Params}). Figure~\ref{fig:pbc_dist} shows the M-PIMC-PBC results for different $V_0$ as a function of the anharmonic contribution. The results corresponding to zero anharmonic contribution represent the PIMC results.
For high barrier height, the particle is trapped close to the potential minima and M-PIMC-PBC show a better acceptance ratio and energy autocorrelation time for any choice of harmonic domain. For intermediate barrier height, M-PIMC-PBC behaves similarly to the non-periodic system, showing an improvement in the acceptance ratio and there is an optimal harmonic domain which leads to the lowest autocorrelation time.
For both cases, M-PIMC-PBC and standard PIMC show a similar energy convergence with the number of beads.
These results are expected since, for larger $V_0$, the worldlines with $W = 0$, which are the ones affected by M-PIMC-PBC, behave very similarly to that of a particle trapped inside a potential well. As a result, M-PIMC-PBC improves the sampling of these worldlines and optimizes the acceptance ratio and the energy autocorrelation time by tuning the harmonic domain.
However, for low barrier height, we find that standard PIMC shows a better acceptance ratio and energy autocorrelation time than M-PIMC-PBC.
In this regime, the worldlines with $W=0$ do not behave like particle inside an anharmonic potential because PBC allows, even for a $W=0$ worldline, to have multiple nonzero local windings while keeping the net winding to zero, making M-PIMC-PBC less efficient.

\begin{figure*}[ht]
\centering
\includegraphics[width=1.0\textwidth]{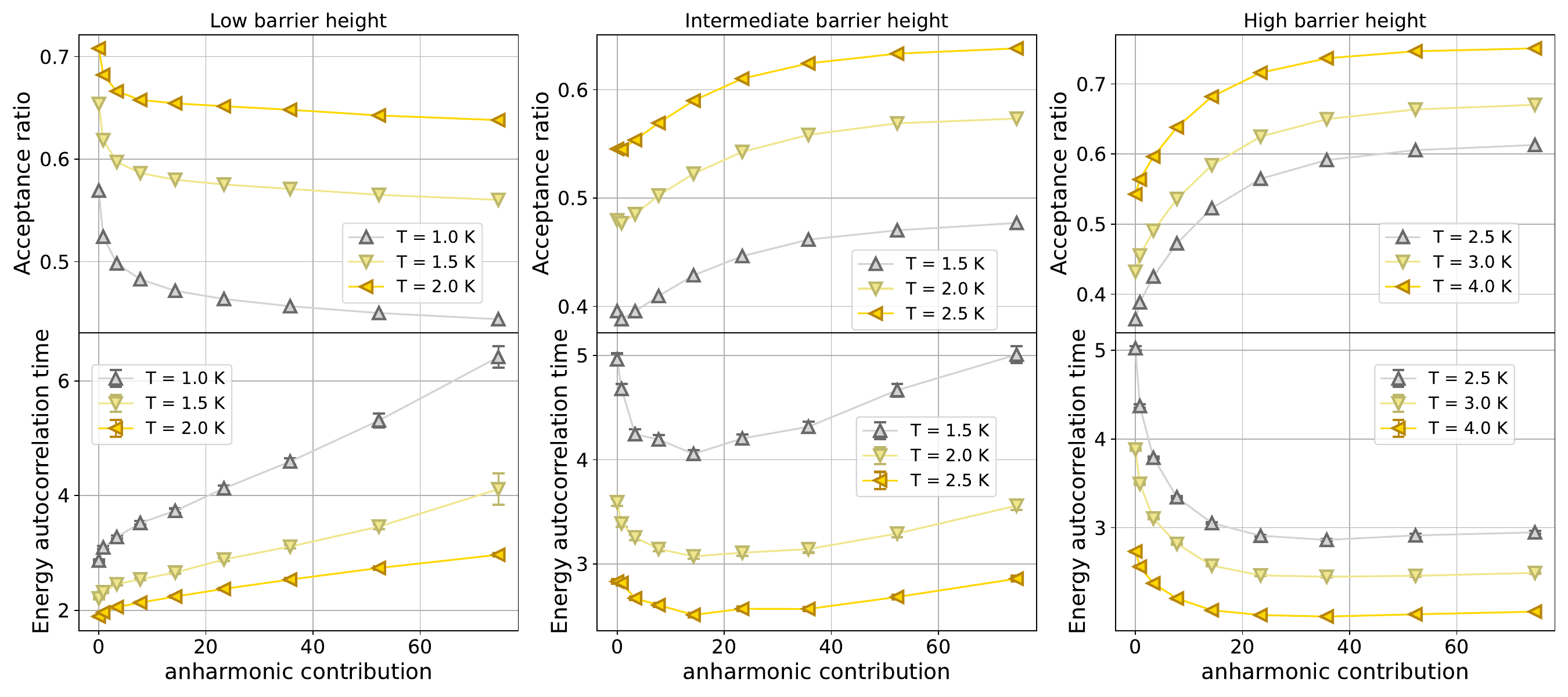}
\caption{M-PIMC-PBC results for a sinusoidal potential $ V(x) = -V_0 \cos\left(\frac{2\pi x}{L}\right)$ are presented for varying barrier height, $V_0$, low barrier height (left column), intermediate barrier height (middle column), and high barrier height (right column), respectively at different temperatures. The points corresponding to zero anharmonic domain represent to the PIMC results. The acceptance ratio and the energy autocorrelation time data are obtained using the number of beads required for the energy convergence at each temperature.}
\label{fig:pbc_dist}
\end{figure*}

\section{Summary and Conclusions}\label{sec: summary}
In this paper, we propose harmonic PIMC (H-PIMC) and its generalization, mixed PIMC (M-PIMC) to improve the sampling in PIMC simulations.  In H-PIMC, trial paths are generated by sampling the harmonic oscillator density matrix exactly and accepting or rejecting them based on the anharmonic potential. H-PIMC is highly efficient for weakly and moderately anharmonic systems improving the acceptance ratio by a factor of 6--16 and reducing the autocorrelation time by a factor of 7--30 at the lowest temperature considered here ($\beta \hbar \omega = 16$). It also leads to faster energy convergence with fewer beads. The faster energy convergence is due to the better energy estimator (Eq.~\ref{eq:Energy_hpimc}) obtained from harmonic Trotter splitting. These advantages become more pronounced as the temperature decreases. However, H-PIMC loses its advantage for strongly anharmonic systems. M-PIMC provides an alternative to H-PIMC in such cases. The M-PIMC algorithm involves identifying a harmonic domain (a tunable parameter) around the local minimum and proposing the harmonic trial paths only there and standard PIMC paths elsewhere. M-PIMC retains all the advantages of H-PIMC while allowing optimization of the energy autocorrelation time for strongly anharmonic systems by tuning the harmonic domain. We also introduce M-PIMC-PBC, an extension of M-PIMC to periodic systems. We benchmark M-PIMC-PBC for a sinusoidal potential with different barrier heights. For high barrier heights, M-PIMC-PBC also shows an optimal harmonic domain with a better acceptance ratio and energy autocorrelation time. Finally, we combine these methods with the worm algorithm to speedup simulations of indistinguishable particles as well. In the future, these methods will accelerate simulations of strongly confined quantum liquids and improve sampling of quantum solids in the presence of defects.

\acknowledgments
B. H. acknowledges support from the Israel Science Foundation (grants No. 1037/22 and 1312/22). S.K. acknowledges support from the Center for Computational Molecular and Materials Science at Tel Aviv University.
S.~P, and A.~D., acknowledge support from the U.S. Department of Energy, Office of Science, Office of Basic Energy Sciences, under Award Number DE-SC0024333. We thank Jacob Higer and Tommer Keidar for useful discussions.

\newpage
\appendix

\section{Details of the parameters}
\label{app:Params}
\subsection{Harmonic potential}
The harmonic potential is
\begin{equation}
    V(x) = \frac{1}{2} m \omega^2 x ^2
\end{equation}
where $m = 1$  atomic unit and $\hbar \omega = 3\ \text{meV} \approx 1.1025 \times 10^{-4}\ $ Hartree.

\subsection{Anharmonic potential}
The anharmonic potential of interest is of the form
\begin{equation}
     V(x) = \frac{1}{2} m \omega^2 {x}^2 \left(1 + c_3 {x} + c_4 {x}^2 \right)
\end{equation}
where $m = 1$  atomic unit, $\hbar \omega = 3\ \text{meV} \approx 1.1025 \times 10^{-4}\ $ Hartree, $c_4 = 10^{-5}\ \text{Bohr}^{-2}$. The different anharmonic regimes considered are
\begin{itemize}
  \item Weak anharmonicity $\Rightarrow$ $c_3 = 0.0025$ Bohr$^{-1}$,
  \item Moderate anharmonicity $\Rightarrow$ $c_3 = 0.0045$ Bohr$^{-1}$,
  \item Strong anharmonicity $\Rightarrow$ $c_3 = 0.0055$ Bohr$^{-1}$.
\end{itemize}
\subsection{Sinusoidal potential}
The sinusoidal potential is of the form
\begin{equation}
    V(x) = -V_0 \cos{\Big(\frac{2\pi x}{L}\Big)}.
\end{equation}
The mass of the particle is, $m = 4$ amu $\approx 7291.5$ atomic unit. The length of the box is, $L = \SI{3.85}{\angstrom} \approx 7.27$ Bohr. The different barrier regimes are,
\begin{itemize}
  \item Low barrier $\Rightarrow$ $V_0 = \SI{0.3}{\milli\eV} \approx 1.1025 \times 10^{-5}\ $ Hartree,
  \item Intermediate barrier $\Rightarrow$ $V_0 = \SI{0.6}{\milli\eV} \approx 2.205 \times 10^{-5}\ $ Hartree,
  \item High barrier $\Rightarrow$ $V_0 = \SI{1.0}{\milli\eV} \approx 3.67 \times 10^{-5}\ $ Hartree.
\end{itemize}

\subsubsection{Low barrier regime (in harmonic oscillator unit)}
The harmonic approximation at the minima results in $\hbar \omega \approx 3.36 \times 10^{-5}\ $ Hartree, where $\omega$ is the corresponding harmonic frequency, and $x_{\text{ho}} = \sqrt{\frac{\hbar}{m\omega}} \approx 2.02$ Bohr. The system parameters in harmonic oscillator unit are
\begin{itemize}
    \item $\tilde V_0 = \frac{V_0}{\hbar \omega} \approx 0.33$
    \item $\tilde L = \frac{L}{x_{\text{ho}}} \approx 3.6$
\end{itemize}

\subsubsection{Intermediate barrier regime (in harmonic oscillator unit)}
The harmonic approximation at the minima results in $\hbar \omega \approx 4.75 \times 10^{-5}\ $ Hartree.,  where $\omega$ is the corresponding harmonic frequency, and $x_{\text{ho}} = \sqrt{\frac{\hbar}{m\omega}} \approx 1.7$ Bohr. The system parameters in harmonic oscillator unit are
\begin{itemize}
    \item $\tilde V_0 = \frac{V_0}{\hbar \omega} \approx 0.46$
    \item $\tilde L = \frac{L}{x_{\text{ho}}} \approx 4.3$
\end{itemize}

\subsubsection{High barrier regime (in harmonic oscillator unit)}
The harmonic approximation at the minima results in $\hbar \omega \approx 6.13 \times 10^{-5}\ $ Hartree,  where $\omega$ is the corresponding harmonic frequency, and $x_{\text{ho}} = \sqrt{\frac{\hbar}{m\omega}} \approx 1.5$ Bohr. The system parameters in harmonic oscillator unit are
\begin{itemize}
    \item $\tilde V_0 = \frac{V_0}{\hbar \omega} \approx 0.6$
    \item $\tilde L = \frac{L}{x_{\text{ho}}} \approx 4.9$
\end{itemize}

% ----------------------------------------------------------------------------------
\section{Detailed balance and generalized acceptance probability for M-PIMC}
\label{app:MPIMC_Detailed_Balance}
% ----------------------------------------------------------------------------------
Consider a staging move with fixed ends $x_{i}$ and $x_{f}$, separated by imaginary time 2$\tau$. Using the staging move, we update the intermediate position $x$ to $y$, i.e., the Monte Carlo move proposes to update the configuration $\textbf{X} \equiv \{x_i, x, x_f \}$ to the configuration $\textbf{Y} \equiv \{x_i, y, x_f \}$.
\begin{equation*}
    \textbf{X}:\{x_i, x, x_f \} \rightarrow \textbf{Y}:\{x_i, y, x_f \}
\end{equation*}
The probability of a configuration \textbf{X} is denoted as $\Pi(\textbf{X}) \equiv \Pi(\{x_i, x, x_f \})$. Under the primitive approximation and using the harmonic Trotter splitting, the probability can be expressed as
\begin{align*}
    \Pi(\{x_i, x, x_f \}) &\approx e^{-\frac{\tau}{2}V_{\text{anh}}(x_i)} \rho_{\text{ho}}(x_i, x;\tau) e^{-\tau V_{\text{anh}}(x)} \\ 
    &\times \rho_{\text{ho}}(x, x_f;\tau) e^{-\frac{\tau}{2}V_{\text{anh}}(x_f)}
\end{align*}
The detailed balance equation corresponding to the move is
\begin{equation*}
    \Pi(\textbf{X}) T(\textbf{Y}|\textbf{X}) A(\textbf{Y}|\textbf{X}) = \Pi(\textbf{Y}) T(\textbf{X}|\textbf{Y}) A(\textbf{X}|\textbf{Y})
\end{equation*}
where, $T(\textbf{Y}|\textbf{X}) \equiv T(\textbf{X} \rightarrow \textbf{Y})$ and $A(\textbf{Y}|\textbf{X}) \equiv A(\textbf{X} \rightarrow \textbf{Y})$ are respectively the trial probability and the acceptance probability for the move $\textbf{X} \rightarrow \textbf{Y}$. Since for this particular move only one bead position is updated, we use the following notation $T(x|y) \equiv T(\textbf{X}|\textbf{Y})$ and $T(y|x) \equiv T(\textbf{Y}|\textbf{X})$ hereafter. Under primitive approximation, we can write the detailed balanced equation as

\begin{widetext}
\begin{align}
   \Rightarrow &e^{-\frac{\tau}{2} V_{\text{anh}}(x_{i})}e^{-\tau V_{\text{anh}}(x)}e^{-\frac{\tau}{2} V_{\text{anh}}(x_{f})}
    \underbrace{\rho_{\text{ho}}(x_{i},x;\tau)\rho_{\text{ho}}(x,x_{f};\tau)}_{\propto T_{\text{ho}}(x|y)}
    T(y|x)A(y|x) \notag \\
    =&\ e^{-\frac{\tau}{2} V_{\text{anh}}(x_{i})}e^{-\tau V_{\text{anh}}(y)}e^{-\frac{\tau}{2} V_{\text{anh}}(x_{f})}
    \underbrace{\rho_{\text{ho}}(x_{i},y;\tau)\rho_{\text{ho}}(y,x_{f};\tau)}_{\propto T_{\text{ho}}(y|x)}
    T(x|y)A(x|y) \notag \\
   \Rightarrow &e^{-\tau V_{\text{anh}}(x)} T_{\text{ho}}(x|y) T(y|x) A(y|x) = e^{-\tau V_{\text{anh}}(y)} T_{\text{ho}}(y|x) T(x|y) A(x|y) \notag \\
   \Rightarrow &A(y|x) = \text{min} \Bigg[1, \frac{T(x|y) / T_{\text{ho}}(x|y)}{T(y|x) / T_{\text{ho}}(y|x)} \frac{e^{-\tau V_{\text{anh}}(y)}}{e^{-\tau V_{\text{anh}}(x)}} \Bigg]
\end{align}
\end{widetext}
In case of the free particle Trotter splitting, it can be shown that
\begin{equation}
    A(y|x) = \text{min} \Bigg[1, \frac{T(x|y) / T_{0}(x|y)}{T(y|x) / T_{0}(y|x)} \frac{e^{-\tau V(y)}}{e^{-\tau V(x)}} \Bigg]
\end{equation}

% ----------------------------------------------------------------------------------
\section{M-PIMC acceptance ratio at zero harmonic domain limit with harmonic Trotter splitting}
\label{app:MPIMC_Acceptance_Ratio}
% ----------------------------------------------------------------------------------
The harmonic oscillator density matrix at inverse temperature $\tau$ is
\begin{widetext}
\begin{align}
\label{eq: rho_ho}
\rho_{\text{ho}}(x',x;\tau) &=\sqrt{\frac{m\omega}{2\pi\hbar\,\sinh(\tau\hbar\omega)}}
\exp\!\left[-\frac{m\omega}{2\hbar\,\sinh(\tau\hbar\omega)} \Big((x'^2+x^2)\cosh(\tau\hbar\omega)-2x'x\Big) \right].
\end{align}

For $\tau \rightarrow 0$, $\tau\hbar\omega \ll 1$. Then,
\begin{equation*}
\sinh\tau\hbar\omega=\tau\hbar\omega+\frac{(\tau\hbar\omega)^3}{6}+O(\tau^5),
\quad
\cosh\tau\hbar\omega=1+\frac{(\tau\hbar\omega)^2}{2}+O(\tau^4).
\end{equation*}

The prefactor in Eq.~\ref{eq: rho_ho}:
\begin{align*}
\sqrt{\frac{m\omega}{2\pi\hbar\,\sinh(\tau\hbar\omega)}} &= \sqrt{\frac{m\omega}{2\pi \hbar \big(\tau \hbar \omega + \frac{(\tau \hbar \omega)^3}{6} + O(\tau^5)\big)}} \\
&= \sqrt{\frac{m}{2\pi\hbar^2\,\tau}} \Big(1-\frac{(\tau\hbar\omega)^2} {12}+O(\tau^4)\Big).
\end{align*}

The exponent in Eq.~\ref{eq: rho_ho}:
\begin{align*}
\frac{m\omega}{2\hbar\,\sinh(\tau\hbar\omega)}
&= \frac{m}{2\hbar^2\tau} - \frac{m\omega^2\,\tau}{12}+O(\tau^3), \\
(x'^2+x^2)\cosh(\tau\hbar\omega)-2x'x &= (x'^2 + x^2)\Big( 1 + \frac{(\tau \hbar \omega)^2}{2} + O(\tau^4) \Big) - 2x'x \notag \\
&= (x'-x)^2 + \frac{(\tau\hbar\omega)^2}{2}(x'^2+x^2)+O(\tau^4).
\end{align*}

Hence,
\begin{align*}
&-\frac{m\omega}{2\hbar\,\sinh(\tau\hbar\omega)} \Big((x'^2+x^2)\cosh(\tau\hbar\omega)-2x'x\Big)\\
&= -\Big(\frac{m}{2\hbar^2\tau} - \frac{m\omega^2\,\tau}{12}+O(\tau^3) \Big) \Big( (x'-x)^2 + \frac{(\tau\hbar\omega)^2}{2}(x'^2+x^2)+O(\tau^4) \Big) \\
&= -\frac{m}{2\hbar^2\tau}(x'-x)^2
-\frac{m\omega^2\,\tau}{4}(x'^2+x^2)
+\frac{m\omega^2\,\tau}{12}(x'-x)^2
+O(\tau^3).
\end{align*}

Thus, in the limit $\tau \rightarrow 0$, Eq.~\ref{eq: rho_ho} becomes
\begin{align*}
\rho_{\text{ho}}(x',x;\tau)
&= \sqrt{\frac{m}{2\pi\hbar^2\tau}}\,
\exp\!\bigg[-\frac{m(x'-x)^2}{2\hbar^2\tau}\bigg] \nonumber \\
&\quad\times \exp\!\bigg[-\frac{\tau}{2}\Big(V_{\text{ho}}(x')+V_{\text{ho}}(x)\Big)\bigg]
\times \exp\!\bigg[\frac{m\omega^2\,\tau}{12}(x'-x)^2\bigg]
\bigg[1-\frac{(\tau\hbar\omega)^2}{12}+O(\tau^4)\bigg],
\end{align*}
with $V_{\text{ho}}(x)=\tfrac12 m\omega^2 x^2$. Now,

\begin{align*}
\rho_{\text{ho}}(x_1,x;\tau)\,\rho_{\text{ho}}(x,x_2;\tau)
&= \frac{m}{2\pi\hbar^2\tau}\,
\exp\!\bigg[-\frac{m}{2\hbar^2\tau}\Big((x_1-x)^2+(x-x_2)^2\Big)\bigg] \nonumber\\
&\quad\times \exp\!\bigg[-\frac{\tau}{2}V_{\text{ho}}(x_1)-\tau V_{\text{ho}}(x)-\frac{\tau}{2}V_{\text{ho}}(x_2)\bigg] \nonumber\\
&\quad\times \exp\!\bigg[\frac{m\omega^2\,\tau}{12}\Big((x_1-x)^2+(x-x_2)^2\Big)\bigg]
\bigg[1-\frac{(\tau\hbar\omega)^2}{6}+O(\tau^4)\bigg].
\end{align*}

The exponent of the first exponential:
\begin{align*}
    -\frac{m}{2\hbar^2\tau}\Big((x_1-x)^2+(x-x_2)^2\Big) &= -\frac{m}{2\hbar^2\tau} \Big(2x^2 - 2x(x_1 + x_2) + (x_1^2 + x_2^2)\Big) \\
    &= -\frac{m}{\hbar^2\tau} \Big(\big(x - \frac{x_1 + x_2}{2}\big)^2 - \frac{(x_1 + x_2)^2}{4} + \frac{x_1^2 + x_2^2}{2}\Big) \\
    &= -\frac{1}{2 \frac{\hbar^2}{2m} \tau} \Big(\big(x - \bar x_{0}\big)^2 - \frac{(x_1 + x_2)^2}{4} + \frac{x_1^2 + x_2^2}{2}\Big) \\
    &= -\frac{1}{2\sigma^2_{0}} \Big(\big(x - \bar x_{0}\big)^2 - \frac{(x_1 + x_2)^2}{4} + \frac{x_1^2 + x_2^2}{2}\Big)
\end{align*}
where, $\sigma_{0}^2 = \frac{\hbar^2}{2m}\tau$ and $\bar x_{0} = \frac{x_1 + x_2}{2}$.

\subsection*{Detailed balance}
The detailed balance equation with the harmonic Trotter splitting and no harmonic domain is
\begin{align*}
    \rho_{\text{ho}}(x_1,x;\tau)\,\rho_{\text{ho}}(x,x_2;\tau) e^{-\tau V_{\text{anh}}(x)} T_{0}(y|x) A(y|x) &= \rho_{\text{ho}}(x_1,y;\tau)\,\rho_{\text{ho}}(y,x_2;\tau) e^{-\tau V_{\text{anh}}(y)} T_{0}(x|y) A(x|y) \notag \\
    \frac{\rho_{\text{ho}}(x_1,x;\tau)\,\rho_{\text{ho}}(x,x_2;\tau) }{T_{0}(x|y)} e^{-\tau V_{\text{anh}}(x)} A(y|x) &= \frac{\rho_{\text{ho}}(x_1,y;\tau)\,\rho_{\text{ho}}(y,x_2;\tau) }{T_{0}(y|x)} e^{-\tau V_{\text{anh}}(y)} A(x|y) \label{db_equation}
\end{align*}
where $T_{0}(x|y) = \frac{1}{\sqrt{2 \pi \sigma_{0}^2}} \exp\Big[{-\frac{(x - \bar x_{0})^2}{2\sigma_{0}^2}}\Big]$.
Now,
\begin{align*}
    \frac{\rho_{\text{ho}}(x_1,x;\tau)\,\rho_{\text{ho}}(x,x_2;\tau)}{T_{0}(x|y)} &\sim \frac{\exp\!\big[-\frac{(x - \bar x_{0})^2}{2\sigma^2_{0}}\big] \times \exp\!\left[-\tau V_{\text{ho}}(x)\right] \times \exp\!\left[\frac{m\omega^2\,\tau}{12}\Big((x_1-x)^2+(x-x_2)^2\Big)\right]}{\exp\!\big[{-\frac{(x - \bar x_{0})^2}{2\sigma_{0}^2}}\big]} \\
    &\sim \exp\!\left[-\tau V_{\text{ho}}(x)\right] \times \exp\!\left[\frac{m\omega^2\,\tau}{12}\Big((x_1-x)^2+(x-x_2)^2\Big)\right]
\end{align*}
The exponent of the second exponential:
\begin{align*}
    (x_1-x)^2+(x-x_2)^2 &= 2x^2 - 2x(x_1 + x_2) + x_1^2 + x_2^2 \\
    &= 2\Big(x - \frac{x_1 + x_2}{2} \Big)^2 - \Big(\frac{x_1 + x_2}{2}\Big)^2 + x_1^2 + x_2^2 \\
    &\sim 2 \sigma_{0}^2 \\
    &\sim 2\tau ,
\end{align*}
since $\sigma^2_{0} = \frac{\hbar^2}{2m} \tau$. Therefore,
\begin{align*}
    \frac{\rho_{\text{ho}}(x_1,x;\tau)\,\rho_{\text{ho}}(x,x_2;\tau)}{T_{0}(x|y)} &\sim \exp\!\left[-\tau V_{\text{ho}}(x)\right] \times \exp\!\left[\frac{m\omega^2\,\tau^2}{6}\big( \cdots \big)\right]
\end{align*}
Then, Eq.~\ref{db_equation} becomes
\begin{align*}
    e^{-\tau V_{\text{ho}}(x)}  \exp\!\left[\frac{m\omega^2\,\tau^2}{6}\big( \cdots \big)\right] e^{-\tau V_{\text{anh}}(x)} A(y|x) = e^{-\tau V_{\text{ho}}(y)} \exp\!\left[\frac{m\omega^2\,\tau^2}{6}\big( \cdots \big)\right] e^{-\tau V_{\text{anh}}(y)} A(x|y)
\end{align*}
\begin{align}
    A(y|x) \approx \text{min} \Big[1, \frac{e^{-\tau V(y)}}{e^{-\tau V(x)}}\Big]
\end{align}
\end{widetext}

\section{Speedup for N=2 bosons}
\label{app: boson_speedup}
We compute the speedup for different harmonic domains for the case of 2 identical particles. We see the region of maximum speedup is the same as that of the single particle case. Thus, we can use the single particle simulations to determine the best possible split for potential.
\begin{figure}
\centering
\includegraphics[width=0.45\textwidth]{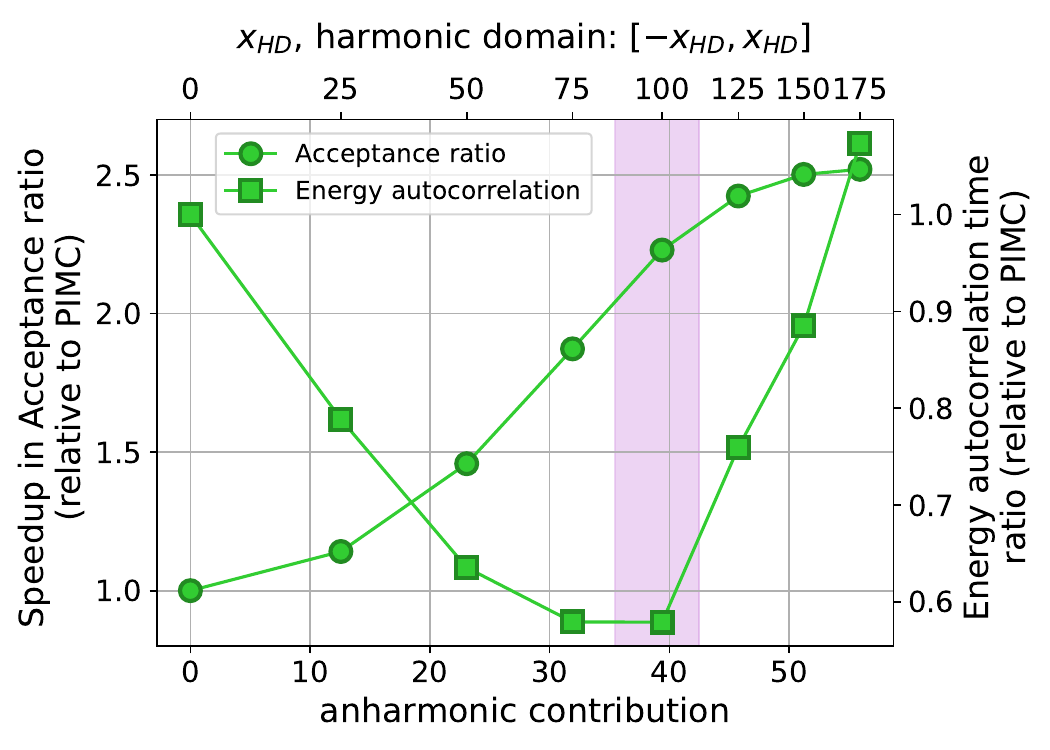} %Placeholder figure
\caption{M-PIMC results for strong anharmonicity at $\beta \hbar \omega = 16$ and N = 2. The M-PIMC calculations use 96 beads. The speedup in acceptance ratio is defined as $\frac{\text{M-PIMC value}}{\text{PIMC value}}$ and the energy autocorrelation time ratio is defined as $\frac{\text{M-PIMC value}}{\text{PIMC value}}$. The shaded area indicates the optimal harmonic domain.}
\label{fig:mpimc_res_2boson}
\end{figure}

\nocite{*}

\bibliography{refs}% Produces the bibliography via BibTeX.

\end{document}